\documentclass[published]{JHEP3} 
\usepackage{amstext}
\usepackage{amsmath}
\usepackage{amssymb}
\usepackage{esint}
\JHEP{00(0000)000}

\JHEPspecialurl{http://jhep.sissa.it/JOURNAL/JHEP3.tar.gz}

\usepackage{epsfig,multicol,bbm}

\newcommand\fverb{\setbox\fverbbox=\hbox\bgroup\verb}
\newcommand\fverbdo{\egroup\medskip\noindent%
			\fbox{\unhbox\fverbbox}\ }
\newcommand\fverbit{\egroup\item[\fbox{\unhbox\fverbbox}]}
\newbox\fverbbox

\title{Logarithmic corrections to black hole and black ring entropy in tunneling
approach}

\author{Yi-Xin Chen\\
	 Zhejiang Institute of Modern Physics, Zhejiang University,
Hangzhou, 310027, China\\
	E-mail: \email{yxchen@zimp.zju.edu.cn}}

\author{Jian-Long Li\\
	 Zhejiang Institute of Modern Physics, Zhejiang University,
Hangzhou, 310027, China\\
	E-mail: \email{marryrene@gmail.com}}

\author{Kai-Nan Shao\\
	 Zhejiang Institute of Modern Physics, Zhejiang University,
Hangzhou, 310027, China\\
	E-mail: \email{shaokn@gmail.com}}

\received{000000} 		
\accepted{000000}		

\preprint{\hepth{0910.5540}}	

\abstract{The tunneling approach beyond semiclassical approximation has been
used to calculate the corrected Hawking temperature and entropy for
various black holes and FRW universe model. We examine their derivations,
and prove that the quantity $H$ in the corrected temperature is the
explicit function of the only free parameter $\mathcal{A}$ (which
is an auxiliary parameter defined by $\mathcal{A}=\hbar S_{BH}$).
Our analysis improves previous calculations, and indicates that the
leading order logarithmic correction to entropy is a natural result
of the corrected temperature and the first law of thermodynamics.
Additionally, we apply the tunneling approach beyond semiclassical
approximation to neutral black rings. Based on the analysis, we show
that the entropy of neutral black rings also has a logarithmic leading
order correction.}


\begin{document} 


\section{Introduction}

Since the discovery of Hawking radiation and black hole thermodynamics\cite{Bardeen1973,Bekenstein1973,Hawking1975},
it is generally believed that black holes are objects which have temperature
and entropy. By adopting the Hawking temperature $T_{H}=\frac{\kappa}{2\pi}$
and the Bekenstein-Hawking entropy $S_{BH}=\frac{A_{H}}{4\hbar}$,
the first law of black hole thermodynamics is obtained\cite{Page2005}.
The understanding of black hole entropy is an important subject in
quantum gravity. When quantum effects are considered, the area law
of black hole entropy should undergo corrections. These corrections
have been extensively studied using many approaches, such as field
theory methods\cite{Fursaev1995,Mann1998,Page2005}, quantum geometry
techniques\cite{Kaul2000,Kloster2008}, general statistical mechanical
arguments\cite{Das2002,More2005,Mukherji2002}, Cardy formula\cite{Carlip2000,Setare2004}.
In all these derivations, the leading order correction to black hole
entropy takes a logarithmic form\begin{equation}
S=S_{BH}+\alpha\ln S_{BH}+\cdots\,\,.\label{eq:LogS}\end{equation}
 This result also coincides with that obtained by counting the number
of microstates in string theory\cite{Strominger1996,Solodukhin1998}
and loop quantum gravity\cite{Rovelli1996,Meissner2004}.

The simplest method of calculating black hole entropy is directly
integrating from the the first law of thermodynamics $dM=TdS$, using
the expression of Hawking temperature. Several derivations of Hawking
radiation were presented\cite{Hartle1976,Gibbons1977,Christensen1977},
one of which is the tunneling approach\cite{Srinivasan1999,Parikh2000}.
In this approach, the emission rate of a particle tunneling from the
black hole is associated with the imaginary part of the action, which
in turn is related to the Boltzmann factor for the emission at the
Hawking temperature. There are two variants of the tunneling approach,
namely, Parikh-Wilczek radial null geodesic method\cite{Parikh2000,Parikh2004},
and the Hamilton-Jacobi method\cite{Srinivasan1999,Shankaranarayanan2001,Shankaranarayanan2002}.
Recently, based on the Hamilton-Jacobi method, Banerjee and Majhi\cite{Banerjee2008,Majhi2009}
developed a general formalism of tunneling beyond semiclassical approximation.
According to their calculations, the Hawking temperature with quantum
corrections takes the form \begin{equation}
T_{cr}=T_{H}\left(1+\sum\gamma_{i}\hbar^{i}\right)=T_{H}\left(1+\sum\beta_{i}\frac{\hbar^{i}}{H^{i}}\right)^{-1}\,\,,\label{eq:T_cr}\end{equation}
 where $H$ is a quantity with the dimension of $\hbar$, and should
be expressed in terms of black hole parameters\cite{Banerjee2009}.
It is reasonable to assume that, the first law of thermodynamics still
holds in the context of quantum corrections, and the entropy $S_{cr}$
is a state function. Then, the entropy can be determined by integrating
the relation $dE=T_{cr}dS_{cr}$, which has a form as (\ref{eq:LogS}).
This formulism has been applied on various black holes\cite{Modak2009,Banerjee2009,Banerjee2009a,Banerjee2009c,Zhu2009a}
and FRW universe model\cite{Zhu2009}. However, after examining their
papers carefully, we find that these derivations rely on one procedure,
i.e., they expressed $H$ as a polynomial of selected black hole parameters.
In fact, all what we know is that $H$ can be expressed in terms of
black hole parameters. There is no reason why $H$ ``should'' be
a polynomial, and why $H$ took those different particular forms in
various models\cite{Modak2009,Banerjee2009,Banerjee2009a,Banerjee2009c,Zhu2009a}.
In order to fill this gap in the calculations of the quantum corrected
entropy using the tunneling approach beyond semiclassical approximation,
we propose to take $H$ as a general function of a complete set of
black hole parameters, i.e. $H=H(\text{black hole parameters})$.
For example, in Kerr-Newman black holes, the complete parameters are
$M$, $J$, and $Q$, while the entropy $S_{BH}$ can be expressed
in terms of them $S_{BH}=S_{BH}(M,Q,J)$. Next we define an auxiliary
$\mathcal{A}$ by $\mathcal{A}:=\mathcal{A}(M,J,Q)=\hbar S_{BH}$.
This is a quantity with area unit, and can stilled be defined via
the above formula even when the entropy $S_{BH}$ does not obey the
Bekenstein-Hawking area law%
\footnote{One example of violating the area law is the Lovelock gravity, for
consistency, in this situation we still use $S_{BH}$ to denote the
standard black hole entropy without quantum correction. Another point
is that one should not confuse $\mathcal{A}$ with the horizon area
of black holes. We shall denote the horizon area by $A_{H}$, if appears
in this paper. %
}. We may then regard the parameters $(J,Q,S_{BH})$, or equivalently
$(J,Q,\mathcal{A})$ as a complete set of parameters for Kerr-Newman
black hole. So $H$ can be expressed as $H=H(M,Q,J)=H(J,Q,\mathcal{A})$.
According to the spirit in \cite{Banerjee2009}, by demanding $dS_{cr}$
to be an exact differential, we can prove that $\frac{\partial H(J,Q,\mathcal{A})}{\partial J}=0=\frac{\partial H(J,Q,\mathcal{A})}{\partial Q}$,
and so $H=H(\mathcal{A})$. Then performing the dimension analysis
as in \cite{Modak2009,Banerjee2009,Banerjee2009a,Banerjee2009c,Zhu2009a},
we can show that the leading order correction to black hole entropy
is logarithmic. Comparing to previous calculations, the treatment
here is more natural and general. In particular, this treatment emphasizes
the use of exact differential in eliminating the extra dependence
in $H$. Another advantage is that the analysis here can be easily
extended to the cases of AdS Schwarzschild black holes, Gauss-Bonnet
and Lovelock black holes, the FRW universe model and neutral black
rings. To this context, we can say that our analysis indicates that
the leading order logarithmic correction to black hole entropy may
be a general result naturally obtained from the corrected temperature
(\ref{eq:T_cr}) and the first law of thermodynamics.

The 5-dimensional black ring is found as a vacuum solution of general
relativity, with the event horizon topology $S^{1}\times S^{2}$\cite{Emparan2006}.
It is very interesting to study the corrected temperature and entropy
of black rings due to quantum effects. The tunneling formulism can
be used to examine the Hawking radiation of black rings\cite{Zhao2007,Jiang2008}.
However, the correction due to quantum effects is not considered.
On the other hand, according to\cite{Chen2008}, the near horizon
dynamics of a scalar field $\varphi$ on black ring metric background
can be reduced to a $(1+1)$ dimensional one. Therefore, the tunneling
formalism beyond semiclassical approximation can also be applied to
black rings. In this Letter, we perform the analysis on neutral black
ring, and show that its corrected temperature and entropy should also
take the form of (\ref{eq:T_cr}) and (\ref{eq:LogS}). As far as
we know, it is the first time to propose that the entropy of neutral
black ring should also undergoes a logarithmic leading order correction
due to quantum effects. In obtaining the corrected entropy from the
temperature (\ref{eq:T_cr}), the previous treatment in \cite{Modak2009,Banerjee2009,Banerjee2009a,Banerjee2009c,Zhu2009a}
seems does not work with black rings. The parameters for neutral black
ring are $R$, $\nu$ and $\lambda$ (with $\lambda=2\nu/(1+\nu^{2})$),
while only $R$ is dimensional. We are not able to determine (or `guess')
a particular polynomial for $H$. However, we may regard $(R,\nu)$,
$(M,J)$ or equivalently $(J,\mathcal{A})$ as a complete set of parameters
for neutral black rings. Then our treatment for Kerr-Newman can easily
be extended on the case of black rings. Although the black hole and
black ring entropy can only be fully determined by counting the microstate
in well defined quantum gravity theories, the study of corrected entropy
due to quantum effects would be helpful for the understanding of these
black objects.

The paper is organized as follows. In Section \ref{sec:2}, we prove
that in the quantity $H$ in the tunneling formalism beyond semiclassical
approximation is an explicit function of the only parameter $\mathcal{A}$.
We illustrate our proof in the context of Kerr-Newman black holes,
and show that this analysis can easily be extended to the cases of
AdS Schwarzschild black holes, Gauss-Bonnet and Lovelock black holes,
and FRW universe model. In Section \ref{sec:3}, we apply the tunneling
formulism on neutral black ring, and obtain the corrected temperature
and entropy. We use the units $G=c=k_{B}=\frac{1}{4\pi\epsilon_{0}}=1$,
while keeping the Planck constant $\hbar$ explicitly. In fact, the
Planck constant is an indication of quantum effects, which can be
used to identify the leading order correction.

\section{Exact differential and logarithmic correction to black hole entropy\label{sec:2}}

In this section, we take the Kerr-Newman black hole as an example
to demonstrate that the leading order correction to the entropy is
logarithmic, and then extend our treatment to other models. Before
calculating the entropy correction from the corrected Hawking temperature
(\ref{eq:T_cr}) obtained in \cite{Banerjee2008,Majhi2009}, we give
a brief review of the first law of thermodynamics on Kerr-Newman black
holes. The standard first law holds on the event horizon \begin{equation}
dS_{BH}=\frac{dM}{T_{H}}+\left(-\frac{\Omega_{H}}{T_{H}}\right)dJ+\left(-\frac{\Phi_{H}}{T_{H}}\right)dQ\,\,,\label{eq:std_1stlaw}\end{equation}
 where $\Omega_{H}$ and $\Phi_{H}$ are the angular velocity and
the electrical potential on the horizon. In Eq.(\ref{eq:std_1stlaw}),
the black hole can be treated as an equilibrium state parametrized
by $M$, $J$ and $Q$, and the entropy $S_{BH}$ is a state function.
Integrate Eq.(\ref{eq:std_1stlaw}) by $M$, $J$ and $Q$, $S_{BH}$
can be expressed as a function of these quantities. For example, the
Bekenstein-Hawking entropy of Kerr-Newman black hole takes the form\begin{equation}
S_{BH}=\frac{\pi}{\hbar}\left(2M\left[M+\sqrt{M^{2}-\frac{J^{2}}{M^{2}}-Q^{2}}\right]-Q^{2}\right)\,\,.\label{eq:S_BH_KN}\end{equation}
 The thermodynamical quantities $T_{H}$, $\Omega_{H}$ and $\Phi_{H}$
can be defined as partial derivatives of $S_{BH}$
\begin{align}
\frac{\partial S_{BH}}{\partial M}\biggr|_{J,Q} & =\frac{1}{T_{H}}\,\,,\label{eq:STH}\\
\frac{\partial S_{BH}}{\partial J}\biggr|_{M,Q} & =-\frac{\Omega_{H}}{T_{H}}\,\,,\label{eq:SOmega}\\
\frac{\partial S_{BH}}{\partial Q}\biggr|_{M,J} & =-\frac{\Phi_{H}}{T_{H}}\,\,.\label{eq:SPhi}\end{align}
 Besides, $S_{BH}$ is a state function and $dS_{BH}$ is an exact
differential. This leads to the following Maxwell's relations\begin{align}
\frac{\partial}{\partial J}\left(\frac{1}{T_{H}}\right)\biggr|_{M,Q} & =\frac{\partial}{\partial M}\left(-\frac{\Omega_{H}}{T_{H}}\right)\biggr|_{J,Q}\,\,,\label{eq:MaxJM}\\
\frac{\partial}{\partial Q}\left(-\frac{\Omega_{H}}{T_{H}}\right)\biggr|_{M,J} & =\frac{\partial}{\partial J}\left(-\frac{\Phi_{H}}{T_{H}}\right)\biggr|_{M,Q}\,\,,\label{eq:MaxQJ}\\
\frac{\partial}{\partial M}\left(-\frac{\Phi_{H}}{T_{H}}\right)\biggr|_{J,Q} & =\frac{\partial}{\partial Q}\left(\frac{1}{T_{H}}\right)\biggr|_{J,M}\,\,.\label{eq:MaxMQ}\end{align}
 When $\hbar\rightarrow0$, the corrected temperature $T_{cr}\rightarrow T_{H}$,
so it is reasonable to assume that the first law of thermodynamics
still holds in the context of quantum correction\cite{Banerjee2009},
i.e.\[
dS_{cr}=\frac{dM}{T_{cr}}+\left(-\frac{\Omega_{H}}{T_{cr}}\right)dJ+\left(-\frac{\Phi_{H}}{T_{cr}}\right)dQ\,\,.\]

The corrected Hawking temperature calculated from the tunneling formalism
beyond the semiclassical approximation is \begin{equation}
T_{cr}=T_{H}\left(1+\sum\gamma_{i}\hbar^{i}\right)=T_{H}\left(1+\sum\beta_{i}\frac{\hbar^{i}}{H_{KN}^{i}}\right)^{-1}\,\,,\label{eq:T_cr_KN}\end{equation}
 where $H_{KN}$ is a quantity with dimension $\hbar$. Similar Maxwell's
relations hold by replacing $T_{BH}$ with $T_{cr}$. By substituting
the expression of corrected temperature (\ref{eq:T_cr_KN}), we obtain
the corresponding constrain equations for $H_{KN}$\begin{align}
\frac{\partial H_{KN}}{\partial J}\biggr|_{M,Q} & =-\Omega_{H}\frac{\partial H_{KN}}{\partial M}\biggr|_{J,Q}\,\,,\label{eq:Cons_JM}\\
\frac{\partial H_{KN}}{\partial Q}\biggr|_{M,J} & =\left(\frac{\Phi_{H}}{\Omega_{H}}\right)\frac{\partial H_{KN}}{\partial J}\biggr|_{M,Q}\,\,,\label{eq:Cons_QJ}\\
\frac{\partial H_{KN}}{\partial M}\biggr|_{J,Q} & =-\frac{1}{\Phi_{H}}\frac{\partial H_{KN}}{\partial Q}\biggr|_{J,M}\,\,.\label{eq:Cons_MQ}\end{align}
 In Kerr-Newman black hole, the three conserved quantities $M$, $J$,
and $Q$ have totally determined the black hole thermodynamics, so
the quantity $H_{KN}$ should be expressed in terms of these black
hole parameters. The next step in \cite{Banerjee2009} is to express
$H_{KN}$ as a particular polynomial of $M,J,Q$, followed by a tedious
analysis using the first law of thermodynamics to determine the expression
of $H_{KN}$. This treatment is not natural and general , and can
be improved. Here we take the general form $H_{KN}=H_{KN}(M,J,Q)$.
Next we introduce an auxiliary parameter $\mathcal{A}$ defined by
$\mathcal{A}=\hbar S_{BH}$. Then we can take $\mathcal{A}$ and two
of the three conserved quantities $M$, $J$, $Q$ as free parameters,
and construct the third conserved quantity as a function of them.
For example, taking $M=M(\mathcal{A},J,Q)$ and substituting it into
$H_{KN}$, we get\[
H_{KN}=H_{KN}(M,J,Q)=H_{KN}^{\prime}(\mathcal{A}(M,J,Q),J,Q)\,\,.\]
 Here we denote $H_{KN}^{\prime}(\mathcal{A},J,Q)$ again as $H_{KN}(\mathcal{A},J,Q)$
for convenience. Substitute $H_{KN}(\mathcal{A}(M,J,Q),J,Q)$ into
the constrain equation (\ref{eq:Cons_JM}), and use the chain rule
in standard calculus textbooks, \begin{align}
LHS. & =\frac{\partial H_{KN}(\mathcal{A},J,Q)}{\partial\mathcal{A}}\biggr|_{J,Q}\frac{\partial\mathcal{A}}{\partial J}\biggr|_{M,Q}\nonumber \\
 & \,+\frac{\partial H_{KN}(\mathcal{A},J,Q)}{\partial Q}\biggr|_{\mathcal{A},Q}\frac{\partial Q}{\partial J}\biggr|_{M,Q}+\frac{\partial H_{KN}(\mathcal{A},J,Q)}{\partial J}\biggr|_{\mathcal{A},Q}\,\,,\label{eq:LHS}\\
RHS. & =-\Omega_{H}\Biggl(\frac{\partial H_{KN}(\mathcal{A},J,Q)}{\partial\mathcal{A}}\biggr|_{J,Q}\frac{\partial\mathcal{A}}{\partial M}\biggr|_{J,Q}\nonumber \\
 & \,+\frac{\partial H_{KN}(\mathcal{A},J,Q)}{\partial J}\biggr|_{\mathcal{A},Q}\frac{\partial J}{\partial M}\biggr|_{J,Q}+\frac{\partial H_{KN}(\mathcal{A},J,Q)}{\partial Q}\biggr|_{\mathcal{A},J}\frac{\partial Q}{\partial M}\biggr|_{J,Q}\Biggr)\,\,.\label{eq:RHS}\end{align}
 The second term of Eq.(\ref{eq:LHS}), the second and the third term
of Eq.(\ref{eq:RHS}) vanish. In addition, the first term of Eq.(\ref{eq:LHS})
cancels with the first term of Eq.(\ref{eq:RHS}) due to Eq.(\ref{eq:SOmega}).
Thus we obtain one constraint of $H_{KN}$\begin{equation}
\frac{\partial H_{KN}(\mathcal{A},J,Q)}{\partial J}\biggr|_{\mathcal{A},Q}=0\,\,.\end{equation}
 Similarly, the other constraint of $H_{KN}$ can be obtained from
the constraint equation (\ref{eq:Cons_MQ})\begin{equation}
\frac{\partial H_{KN}(\mathcal{A},J,Q)}{\partial Q}\biggr|_{\mathcal{A},J}=0\,\,.\end{equation}
 As a result, once we take $\mathcal{A}$ as a free parameter in describing
$H_{KN}$, it is the only explicit free parameter, i.e. \begin{equation}
H_{KN}=H_{KN}(\mathcal{A})\,\,.\label{eq:HKN_proof}\end{equation}
 If we take $J=J(\mathcal{A},M,Q)$ or $Q=Q(\mathcal{A},M,J)$, Eq.(\ref{eq:HKN_proof})
can also be obtained from the constraint equations (\ref{eq:Cons_JM}-\ref{eq:Cons_MQ}).
This indicates that the result (\ref{eq:HKN_proof}) is a naturally
consequence of the first law of thermodynamics. Now we have completed
the proof that $H_{KN}$ is the function of the only explicit parameter
$\mathcal{A}$.

In D-dimensional spacetime, the units $G=c=k_{B}=\frac{1}{4\pi\epsilon_{0}}=1$,
and$\left[H_{KN}\right]=\left[\hbar\right]=\left[\mathcal{A}\right]=L^{D-2}$
, so in general $H_{KN}$ can be expressed as%
\footnote{One can expect that the dependence of $\hbar$ in $T_{cr}$ has been
expressed in (\ref{eq:T_cr_KN}), so that $\hbar$ does not appear
in the expression of $H_{KN}$. %
} \begin{equation}
H_{KN}=\eta\mathcal{A}\quad,\label{eq:HKN_expand}\end{equation}
 where $\eta$ is a dimensionless parameter. Using the first law of
thermodynamics\begin{align*}
dS_{cr} & =\frac{1}{T_{cr}}\left(dM-\Omega_{H}dJ-\Phi_{H}dQ\right)\\
 & =\frac{1}{T_{cr}}\left(T_{H}dS_{BH}\right)\,\,,\end{align*}
 which can be written as\begin{align}
dS_{cr} & =\left(1+\sum\beta_{i}\frac{\hbar^{i}}{H_{KN}^{i}}\right)dS_{BH}\nonumber \\
 & =\left(1+\beta_{1}\frac{\hbar}{\eta\mathcal{A}}+O(\hbar^{2})\right)dS_{BH}\nonumber \\
 & =\left(1+\beta_{1}\frac{1}{\eta S_{BH}}+O(\hbar^{2})\right)dS_{BH}\nonumber \\
 & =\left(1+\frac{\alpha}{S_{BH}}+\cdots\right)dS_{BH}\label{eq:dS_cr2dS}\end{align}
 where $\alpha$ is another dimensionless coefficient, and all other
higher order terms are included in $(\cdots)$. Integrate Eq.(\ref{eq:dS_cr2dS})
up to the first (leading) order, which yields\begin{equation}
S_{cr}=S_{BH}+\alpha\ln S_{BH}+\cdots+const\,\,\,.\label{eq:LogS_Obtained}\end{equation}
 This is the corrected entropy of Kerr-Newman black hole obtained
from the temperature (\ref{eq:T_cr_KN}) and the leading order correction
is logarithmic.

Our analysis can be easily extended to the cases of various black
holes and FRW universe model\cite{Banerjee2009,Modak2009,Banerjee2009a,Banerjee2009c,Zhu2009a,Zhu2009},
without assigning $H$ as specific polynomials for each case : 
\begin{itemize}
\item For Schwarzschild black hole , the only conserved quantity is mass
$M$, and $S_{BH}=\frac{A_{H}}{4\hbar}=\frac{4\pi M^{2}}{\hbar}\equiv\frac{\mathcal{A}}{\hbar}$
so $H=H(M)=H(\mathcal{A})$. The cases of Kerr black hole and Reissner-Nordstrom
black hole is similar to the Kerr-Newman black hole. 
\item For FRW universe model, the parameter is $\tilde{r}_{A}$, the radius
of apparent horizon, and $S_{BH}=\frac{\pi\tilde{r}_{A}^{2}}{\hbar}\equiv\frac{\mathcal{A}}{\hbar}$.
So $H(\tilde{r}_{A})=H(\mathcal{A})$. 
\item For AdS Schwarzschild black hole, the situation is a bit more complicated.
The mass and entropy of AdS Schwarzschild black hole is \[
M=\frac{r_{+}(1+r_{+}^{2}/l^{2})}{2}\,,\,\,\, S=\frac{\pi r_{+}^{2}}{\hbar}\,\,,\]
 where $\Lambda=-3l^{2}$, $l$ is the AdS radius, and $[l]^{2}=L^{2}=[\hbar]$.
Since the cosmological constant comes into the black hole solution,
it modifies the form of first law of black hole thermodynamics. Thus
it is reasonable to consider an ensemble where the cosmological constant
can fluctuate\cite{Caldarelli2000}, \begin{equation}
dS=\frac{1}{T}(dM-\Omega dJ-\Phi dQ-\Theta d\Lambda).\label{eq:Cosmo}\end{equation}
 Then we can prove that $H=H(\mathcal{A})$. This result not only
preserves the consistency of the black hole thermodynamics in AdS
case, but also ensures that when $\Lambda\rightarrow0$, the quantum
corrections of black hole temperature and entropy degenerate to the
Schwarzschild case. 
\item For D-dimensional Gauss-Bonnet black hole , \[
S_{GB}=\frac{A_{D-2}r_{h}^{D-2}}{4\hbar}\left(1+\frac{D-2}{D-4}\frac{\bar{\lambda}}{r_{h}^{2}}\right)\,\,,\]
 where $\bar{\lambda}=\lambda(D-3)(D-4)$, $[\bar{\lambda}]=L^{2}$,
and $\lambda$ is the coupling constant of the Gauss-Bonnet term (see
\cite{Banerjee2009c}). It undergoes similar situation as the AdS
case, and we can also get $H=H(\mathcal{A})$. 
\end{itemize}
There is one comment on the corrected entropy of AdS Schwarzschild
black hole. In the treatment of \cite{Banerjee2008}, they set $H=M^{2}$
for the AdS Schwarzschild black hole (See Eq.(4.9) and (4.10) therein).
Following their treatment, we can write the terms of quantum corrections
to the entropy explicitly as\begin{equation}
S_{cr}-S_{BH}=\ln S_{BH}+\frac{\pi l^{2}}{\pi l^{2}+\hbar S_{BH}}-\ln\left(\frac{\pi l^{2}}{\hbar}+S_{BH}\right)+\dots\quad.\label{eq:Scr_ref}\end{equation}
When the cosmological constant vanishes, i.e., as $\Lambda=-3l^{-2}\rightarrow0$,
the dynamics of AdS Schwarzschild black hole should reduce to the
corresponding asymptotically flat Schwarzschild black hole. However,
as $\Lambda\rightarrow0$, the above correction (\ref{eq:Scr_ref})
is not logarithmic, which does not coincide with the the corrected
entropy of Schwarzschild black hole. So the treatment in \cite{Banerjee2008}
cannot be correct. After examining their derivation, we think that
the choice of $H=M^{2}$ for AdS Schwarzschild black hole is groundless
and dogmatic. Their result in Gauss-Bonnet black hole suffers similar
problems. In our paper,we can overcome this arbitrariness, and obtain
the corrected leading order logarithmic correction to black hole entropy.

Now we have examined the derivations of corrected entropy in the cases
of these various black holes \cite{Modak2009,Banerjee2009,Banerjee2009a,Banerjee2009c,Zhu2009a},
FRW universe model\cite{Zhu2009}, and added the proof of the expression
$H=H(\mathcal{A})$. Our treatment strengths the validity of the method
of obtaining the corrected entropy from the tunneling formalism beyond
semiclassical approximation. The analysis indicate that the leading
order entropy correction is a natural result obtained from the corrected
temperature (\ref{eq:T_cr_KN}) and the first law of thermodynamics.
In the next section we shall extend the above analysis to neutral
black rings, and shows that its leading order entropy correction due
to quantum effects is also logarithmic.

\section{Corrected entropy of neutral black rings\label{sec:3}}

The 5-dimensional black ring is a vacuum solution of general relativity,
with the event horizon topology $S^{1}\times S^{2}$. The tunneling
approach has also been used to analyze the black ring radiation\cite{Zhao2007,Jiang2008}.
However, the influence of quantum effects are not included. In this
section we apply the tunneling method beyond semiclassical approximation
on neutral black rings. The metric of neutral black ring is\begin{align*}
ds^{2} & =-\frac{F(y)}{F(x)}(dt-CR\frac{1+y}{F(y)}d\psi)^{2}+\frac{R^{2}}{(x-y)^{2}}F(x)\\
 & \times\biggl[-\frac{G(y)}{F(y)}d\psi^{2}-\frac{dy^{2}}{G(y)}+\frac{dx^{2}}{G(x)}+\frac{G(x)}{F(x)}d\phi^{2}\biggr]\end{align*}
 where $F(\xi)=1+\lambda\xi$ ,$G(\xi)=(1-\xi^{2})(1+\nu\xi)$, $C(\nu,\lambda)=\sqrt{\lambda(\lambda-\nu)\frac{1+\lambda}{1-\lambda}}$.
$\lambda$, $\nu$ are dimensionless parameters taking values in the
range $0<\nu\leqslant\lambda<1$. $x$ is one of the angular coordinates
in $S^{2}$, with $-1\leqslant x\leqslant1$. $y$ is only {}``ringlike-radical''
coordinate, with $-\infty\leqslant y\leqslant-1$. To remove the conical
singularity at $x=1$, $\lambda$ and $\nu$ must be related to each
other via\begin{equation}
\lambda=\frac{2\nu}{1+\nu^{2}}\,\,.\label{eq:lamdanu}\end{equation}
 The horizon is at $y=y_{h}=-1/\nu$ with the topology $S^{1}\times S^{2}$.

The thermodynamics quantities of neutral black ring are \begin{align}
M & =\frac{3\pi R^{2}}{4}\frac{\lambda}{1-\nu}\,,\label{eq:M-br}\\
J & =\frac{\pi R^{3}}{2}\frac{\sqrt{\lambda(\lambda-\nu)(1+\lambda)}}{(1-\nu)^{2}}\,,\label{eq:J-br}\\
S_{\mathrm{BR}} & =\frac{{\cal A}_{h}}{4\hbar}=\frac{2\pi^{2}R^{3}}{\hbar}\frac{\nu^{3/2}\sqrt{\lambda(1-\lambda^{2})}}{(1-\nu)^{2}(1+\nu)}\,,\label{eq:S-br}\\
T_{\mathrm{BR}} & =\frac{\hbar}{4\pi R}(1+\nu)\sqrt{\frac{1-\lambda}{\lambda\nu(1+\lambda)}}\,,\label{eq:T-br}\\
\Omega_{\mathrm{h}} & =\frac{1}{R}\sqrt{\frac{\lambda-\nu}{\lambda(1+\lambda)}}\,.\label{eq:Omg-br}\end{align}
 Here $R$ describes the scale for the solution, and $\lambda$ and
$\nu$ are parameters that determine the shape and rotation velocity
of the ring\cite{Emparan2006}. The validity of first law $TdS=dM-\Omega_{h}dJ$
also requires that $\lambda$ and $\nu$ satisfy the relation (\ref{eq:lamdanu}).
As a result, a neutral black ring parametrized by $M$ and $J$ can
be equivalently described by $R$ and $\nu$. Furthermore, according
to \cite{Chen2008}, a scalar field $\varphi$ in the neutral black
ring reduces to a (1+1)-dimensional free field in the near horizon
limit. So the tunneling formalism beyond semiclassical approximation
can also be applied to neutral black ring. Now let us turn to the
(1+1)-dimensional metric\begin{equation}
ds^{2}=-f(y)dt^{2}+\frac{1}{f(y)}dy^{2}\,\,,\label{eq:br11d}\end{equation}
 where, \[
f(y)=\frac{\sqrt{F(y)}}{CR(1+y)}G(y)\,\,.\]
 Following the standard treatment, consider a massless particle in
spacetime (\ref{eq:br11d}) described by the Klein-Gordon equation\begin{equation}
-\frac{\hbar^{2}}{\sqrt{-g}}\partial_{\mu}\left[g^{\mu\nu}\sqrt{-g}\partial_{\nu}\right]\phi=0\,\,.\label{eq:brKG}\end{equation}
 The semiclassical wave function satisfying the above equation is
obtained by making the standard WKB ansatz for $\phi$\begin{equation}
\phi(y,t)=\exp\left[-\frac{i}{\hbar}\mathcal{S}(y,t)\right]\,\,.\label{eq:brWKB}\end{equation}
 Substitute it in (\ref{eq:brKG}), and expanding the action $\mathcal{S}(y,t)$
in the powers of $\hbar$ \[
\mathcal{S}(y,t)=\mathcal{S}_{0}(y,t)+\sum_{i}\hbar^{i}\mathcal{S}_{i}(y,t)\,\,.\]
 Following the similar treatment of Kerr-Newman black hole in \cite{Banerjee2009},
the final result of the corrected Hawking temperature is \begin{equation}
T_{cr}=T_{BR}\left(1+\sum\beta_{i}\left(\frac{\hbar}{H_{BR}}\right)^{i}\right)^{-1}\,\,.\label{eq:T_cr_BR}\end{equation}
 Derivation of (\ref{eq:T_cr_BR}) via fermion tunneling is also available,
similar to the treatment of other black hole models.

In the context, we have stated that the thermodynamics of neutral
black ring can be equivalently parametrized by $M$ and $J$, or by
$R$ and $\nu$. The neutral black ring satisfies the Bekenstein-Hawking
law, therefore $\mathcal{A}=\frac{A_{h}}{4}$. Following the analysis
in the previous section, we can prove that $H_{br}=H(R,\nu)=H(\mathcal{A})$.
Finally, integrating an equation similar as (\ref{eq:dS_cr2dS}),
the corrected entropy can be obtained \begin{equation}
S_{cr}=S_{BR}+\alpha\ln S_{BR}+\cdots+const\,\,.\label{eq:logS_br}\end{equation}
 Now based on the tunneling formulism beyond semiclassical approximation
and the first law of thermodynamics, we have shown that the leading
order entropy correction of neutral black ring due to quantum effects
is also a logarithmic term.

\section{Summary\label{sec:4}}

The standard entropy of black holes and black ring entropy should
undergoes corrections due to quantum effects. The tunneling formulism
beyond semiclassical approximation provides a useful method to calculate
the corrected Hawking temperatures and entropy in various models.
The entropy corrections obtained using this method have a logarithmic
leading order term. This is a natural result from the corrected temperature
and the first law of thermodynamics, and coincides with some result
obtained by counting the number of microstates in string theory and
loop quantum gravity. Maybe this is a somewhat universal phenomenon.
In fact, for a black hole model, if the quantum tunneling beyond semiclassical
approximation can be used and the corrected temperature can be expressed
as (\ref{eq:T_cr}), the leading order correction to its entropy should
undergo a logarithmic form, according to the analysis in this paper.
However, the exact expressions of black hole and black ring entropy
can only be determined in a well-defined quantum gravity theory. We
hope that our discussion would be helpful for further understanding
of black hole entropy.

\begin{acknowledgments}

We thank C.Cao, Q.J.Cao, Y.J.Du, and Q.Ma for useful discussions.
The work is supported in part by the NNSF of China Grant No. 90503009,
No. 10775116, and 973 Program Grant No. 2005CB724508.

\end{acknowledgments}

\bibliographystyle{JHEP}
\bibliography{logentropy}

\end{document}